\documentclass{article}
\usepackage{ijcai26}
\usepackage{natbib}
\usepackage{times}
\usepackage{soul}
\usepackage{url}
\usepackage[hidelinks]{hyperref}
\usepackage[utf8]{inputenc}
\usepackage[small]{caption}
\usepackage{graphicx}
\usepackage{graphicx}
\usepackage{booktabs}
\usepackage{caption}
\usepackage{amsmath}
\usepackage{amsthm}
\usepackage{booktabs}
\usepackage{algorithm}
\usepackage{algorithmic}
\usepackage[switch]{lineno}
\usepackage{xcolor}
\usepackage{booktabs}
\usepackage{amsmath,amssymb}
\usepackage{multirow}
\usepackage{threeparttable}
\usepackage{tcolorbox}
\tcbset{
    colback=gray!5,
    colframe=black!60,
    boxrule=0.5pt,
    arc=3pt,
    left=6pt,
    right=6pt,
    top=6pt,
    bottom=6pt,
}

\title{From Hypotheses to Factors: Constrained LLM Agents in Cryptocurrency Markets}

\author{
Yikuan Huang$^{1,2,*}$
\and
Zheqi Fan$^{1,2,*}$
\and
Kaiqi Hu$^{3}$
\and
Yifan Ye$^{4}$
\\
\affiliations
$^1$Division of EMIA,
Hong Kong University of Science and Technology, Hong Kong SAR\\
$^2$Thrust of FinTech,
Hong Kong University of Science and Technology, Guangzhou, China\\
$^3$Rutgers Business School,
Rutgers University–New Brunswick, New Jersey, U.S.A.\\
$^4$Faculty of Business and Management, Beijing Normal–Hong Kong Baptist University, Zhuhai, China\\
\textsuperscript{*}Corresponding authors.
\emails
\{yk.huang, zheqi.fan\}@connect.ust.hk
\and
kaiqi.hu7@rutgers.edu
\and
yifanye@bnbu.edu.cn
}

\begin{document}

\maketitle

\begin{abstract}
LLM agents are promising tools for empirical discovery, but their flexibility
can also turn discovery into uncontrolled search. We study how to use agents
under a reproducible protocol through cryptocurrency factor discovery. Our
framework casts the task as sequential hypothesis search: an agent reads an
append-only experiment trace, proposes falsifiable factor hypotheses, and maps
them to executable recipes, while a deterministic engine enforces fixed data
splits, selection gates, transaction costs, and portfolio tests. Candidate
actions are restricted to a point-in-time factor DSL, making both successful and
failed hypotheses auditable. A ridge-combined portfolio
trained only on 2020--2022 data achieves a 44.55\% annualized return and Sharpe
ratio of 1.55 in the 2024--2026 pure out-of-sample period after a 5 basis point one-way trading cost.
\end{abstract}

\section{Introduction}
\label{sec:introduction}

Large language model (LLM) agents can read prior results, propose hypotheses,
write programs, and revise plans from feedback. These abilities are useful for
empirical discovery, but they also create a methodological risk: if an agent can
freely change features, code, data splits, or evaluation rules, strong results
may reflect uncontrolled search rather than valid discovery. This paper asks
whether an agent can discover predictive patterns inside a protocol that is
reproducible, auditable, and robust to data mining.

\begin{figure}[htbp]
\begin{center}
\includegraphics[width=\linewidth]{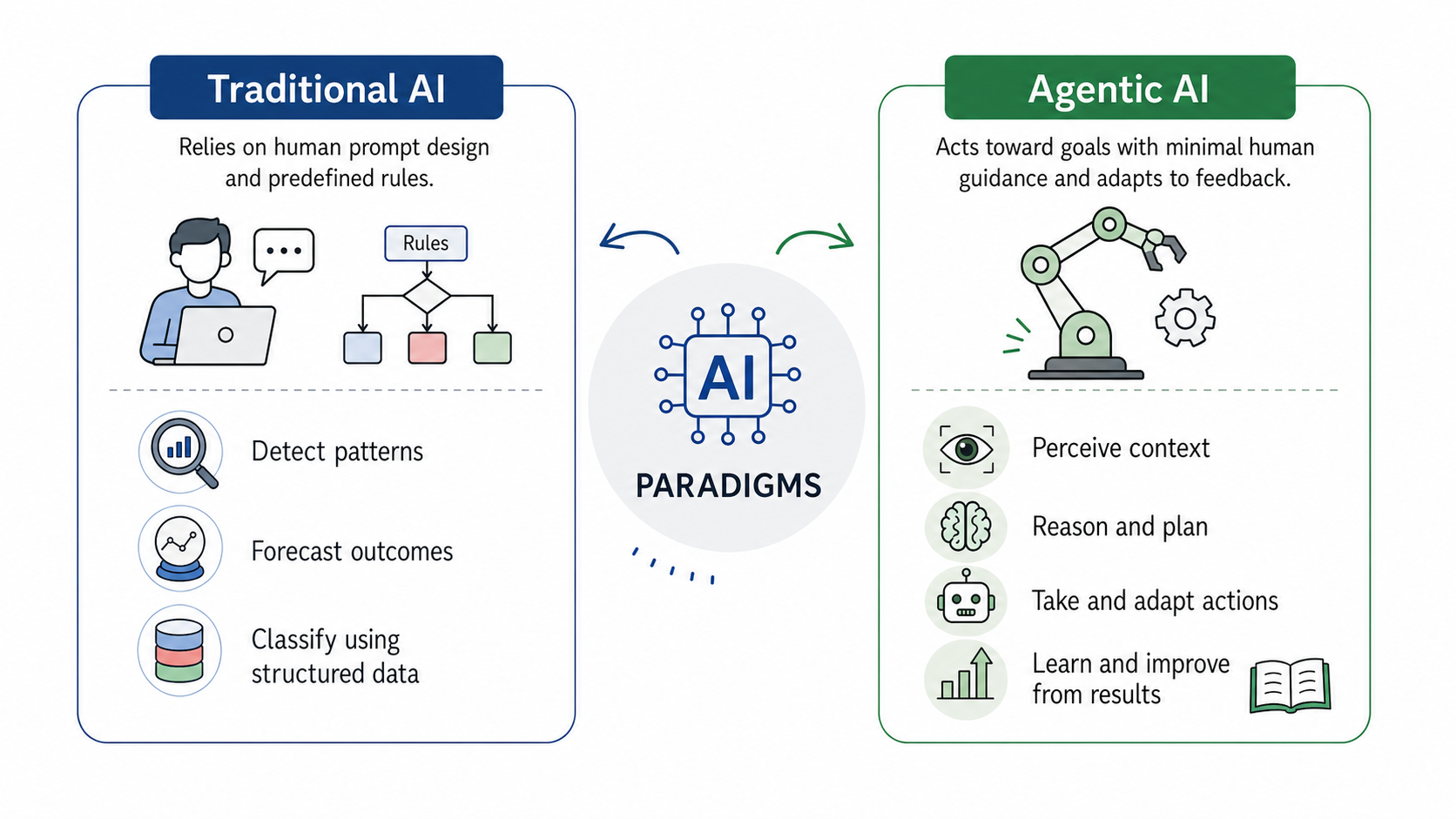}
\caption{Core Capabilities Comparison: Traditional vs. Agentic AI}
\label{fig:agenticAI}
\end{center}
\end{figure}

We study this question through financial factor discovery. Factor research is
not only a supervised prediction problem. A researcher must decide what
mechanism to test, how to express it as a signal, which failures are
informative, and how the next candidate should differ from previous ones. This makes factor discovery a sequential hypothesis search problem, a natural setting
for agentic reasoning if the empirical protocol is kept fixed.
As illustrated in Figure~\ref{fig:agenticAI}, agentic AI differs from traditional AI by operating through goal-directed reasoning, adaptive action, and iterative feedback loops rather than predefined static rules.

We propose an agentic factor discovery framework that separates reasoning from
evaluation. The agent observes an append-only trace of previous candidates and
metrics, proposes falsifiable factor hypotheses, maps them to executable signal
recipes, and interprets the evidence. A deterministic evaluation engine computes
all signals, applies pre-specified selection gates, records failures, and
performs out-of-sample portfolio tests. The agent controls the direction of
search, but it cannot modify the evaluation rule within a search session.

To make the search inspectable, each candidate is written in a constrained
factor domain-specific language (DSL) over point-in-time market variables. The
DSL supports cross-sectional ranks, time-series transforms, nonlinear
transforms, and linear combinations, but prevents arbitrary code generation and
forward-looking features. Each experiment trace records the hypothesis,
rationale, recipe, empirical metrics, interpretation, and next-round decision.
Thus the output is not only a set of factors, but also a reviewable account of
how the agent used positive and negative evidence.

We instantiate the framework in cryptocurrency markets, where heterogeneous
listing histories, sharp liquidity differences, and fast-moving speculative
attention make manual factor search difficult. Our cleaned daily panel covers from January 2020 to December 2025. Candidate selection uses only the
2020--2022 training window, validation results are reported for 2023, and pure
out-of-sample performance is reserved for 2024 onward.

Across five search rounds, the agent converges from broad exploration over
size, volatility, range, liquidity, and momentum toward a compact mechanism:
small, liquidity-scarce tokens with persistent intraday range and positive trend
tend to outperform. Failed hypotheses are also informative: range changes and
volume recoveries are noisy, while level-based scarcity and attention measures
are more stable. A ridge-combined equal-weight long-short portfolio trained only
in sample achieves a 44.55\% annualized return and Sharpe ratio of 1.55 in pure
out-of-sample testing after a 5 basis point one-way trading cost. The
market-cap-weighted version performs poorly, indicating that the alpha is
concentrated in smaller tokens and is capacity constrained.

This paper makes three contributions. First, we formulate factor discovery as
agentic sequential hypothesis search, with symbolic hypotheses as actions and
deterministic empirical tests as feedback. Second, we introduce an auditable
workflow that links natural-language hypotheses to executable signal definitions
through a constrained DSL, fixed gates, and append-only traces. Third, we
provide a cryptocurrency testbed showing that agent-guided search can discover a
coherent family of out-of-sample factors while exposing their limitations
through transaction cost and capacity analyses.

The remainder of the paper is organized as follows.
Section~\ref{sec:literature} reviews the related literature. 
Section~\ref{sec:problem} formalizes the search problem.
Section~\ref{sec:framework} describes the framework. Subsequent sections
present the data, experiments, portfolio tests, and limitations.
Section~\ref{sec:experiments} demonstrates our main empirical results on crypto markets.
Section~\ref{sec:conclusion} concludes the paper.

\section{Related work}
\label{sec:literature}
\noindent Our research bridges financial economics and artificial intelligence. By synthesizing cryptocurrency asset pricing, technical trading signals, and agentic AI, we propose a methodological shift: moving from human-engineered factors toward an autonomous, self-refining AI system for signal discovery and validation.

\subsection{Cryptocurrency Asset Pricing}
\noindent As digital assets matured, the literature on crypto pricing has expanded rapidly. Lacking standard fundamental metrics, researchers have identified crypto-specific drivers of expected returns. Foundational work by \cite{liu2021risks} and \cite{liu2022common} established a three-factor model (market, size, momentum) that captures significant cross-sectional variation. 
Macro-financial states also influence returns; \cite{lee2026state} show that crypto assets are particularly sensitive to credit spread widening, reflecting the fragility of leveraged retail traders.

Given the scarcity of fundamental data, many studies focus on price-volume dynamics. \cite{grobys2020technical} find that simple moving average strategies yield excess returns, suggesting market inefficiencies, while \cite{yang2019behavioral} note that behavior-driven momentum dominates risk-based anomalies. Although crypto momentum is profitable, it carries severe crash risks, necessitating volatility management \citep{grobys2025cryptocurrency}. Recent advances include ``CTREND,'' an aggregated trend factor that outperforms standard models \citep{fieberg2025trend}. Furthermore, comprehensive surveys \citep{baybutt2024empirical,fang2025cryptocurrency} and studies utilizing alternative data and machine learning \citep{maitre2025social, djanga2025cryptocurrency} confirm that microstructure and sentiment signals contain rich predictive content for short-term crypto returns.

\subsection{Factor Models and Machine Learning}
\noindent Empirical asset pricing has transitioned from human-engineered factors \citep{fama2015five, lo2000foundations, chen2025cross} to systematic machine learning (ML) integration. Landmark studies show that ML models, such as deep neural networks, outperform linear regressions by capturing non-linear risk exposures \citep{gu2020empirical}. 
However, standard ML applications face critical bottlenecks. \cite{avramov2023machine} highlight that these models often act as uninterpretable ``black boxes'' susceptible to overfitting. Moreover, while ML optimizes input weights, features are still manually pre-selected, limiting functional forms to human imagination. While genetic programming attempts to bypass this by searching non-differentiable functional spaces \citep{brogaard2023machine}, manual feature reliance remains a limitation. 

Existing financial AI literature largely focuses on predictive modeling for asset returns \citep{gu2020empirical, choi2025alpha}, or structural estimations of pricing models \citep{fan2025options, fan2026deep}. In contrast, we shift toward autonomous strategy engineering. By reframing factor discovery as an LLM-driven AutoML and symbolic regression problem, we navigate the space of mathematical functional forms autonomously. This framework addresses the black-box critique by producing interpretable formulas and mitigating overfitting via transparent selection gates and empirical feedback loops.

\subsection{LLMs and Agentic AI in Finance}
\noindent The integration of Natural Language Processing (NLP) in finance has progressed from basic dictionary-driven sentiment scoring \citep{tetlock2007giving} to leveraging the advanced cognitive capabilities of Large Language Models (LLMs). LLMs have demonstrated remarkable capabilities in financial text analysis, moving beyond simple sentiment analysis to complex reasoning tasks and interpretable internal representations \citep{wu2023bloomberggpt, kong2024large1, xie2024finben}. Unlike traditional methods, LLMs can parse nuanced context, allowing for applications that range from macroeconomic forecasting \citep{chen2025chatgpt} and corporate event analysis \citep{xie2023pixiu} and the tracking of evolving semantic signals in corporate disclosures \citep{choi2025text}.

Building on these foundational reasoning capabilities, researchers have applied LLMs to direct return prediction. \cite{lopez2023can} illustrated that ChatGPT can interpret financial news headlines to predict stock movements, vastly outperforming legacy sentiment metrics. \cite{chen2022expected} utilized LLMs to extract high-dimensional semantic embeddings from news, significantly enhancing cross-sectional return forecasts globally. In this domain, \cite{huang2026cross} leverage LLMs for semantic reasoning to filter spurious correlations in embedding-based financial networks, thereby strengthening cross-stock predictability. 
Moving beyond pure text, researchers are increasingly adapting LLM architectures to process structured financial data. To address the critical issue of lookahead bias, \cite{he2025chronologically} introduced ChronoBERT, ensuring strict chronological data processing for credible backtesting. The inferential power of LLMs has also been harnessed for factor generation; \cite{cheng2024gpt} demonstrated that GPT-4 can independently deduce profitable trading factors.

Most recently, the frontier has shifted toward fully autonomous AI agents. \cite{chen2025agentic} explored how agentic AI is transforming the operational pipelines of asset management, while \cite{huang2026beyond} develops an autonomous framework for systematic factor investing via agentic AI. Our research directly tackles these reliability and hallucination risks by grounding the AI in a strict empirical environment. Rather than using LLMs as passive text analyzers or one-off code generators, we operationalize the AI as an autonomous quantitative researcher within a closed-loop system. The agent proposes logic, executes backtests, and refines its hypotheses based solely on historical crypto market feedback.

\section{Problem Formulation}
\label{sec:problem}

We formulate factor discovery as a sequential hypothesis search problem. The
objective is not to learn a direct mapping from market states to returns, but to
discover a set of interpretable factor definitions that are empirically
predictive, economically meaningful, and robust outside the data used for
selection.

\subsection{Market Panel and Prediction Target}

Let $\mathcal{I}_t$ denote the set of tradable assets available on date $t$.
For each asset $i \in \mathcal{I}_t$, the researcher observes a point-in-time
feature vector
\begin{equation}
    x_{i,t} \in \mathbb{R}^{p},
\end{equation}
constructed only from information available at or before date $t$. In our
empirical setting, $x_{i,t}$ contains daily price, volume, market capitalization,
and derived lagged or rolling variables from a cryptocurrency panel.

The prediction target is the forward return over a fixed execution lag and
holding horizon. With a one-day execution lag and one-day holding period, the
target is
\begin{equation}
    r_{i,t+1:t+2}
    =
    \frac{P_{i,t+2}}{P_{i,t+1}} - 1,
\end{equation}
where $P_{i,t+1}$ is the entry price and $P_{i,t+2}$ is the exit price. The
lag ensures that a signal formed at date $t$ is not evaluated using an
unrealistic same-close execution assumption.

A factor is a scoring function
\begin{equation}
    f_{\theta}: x_{i,t} \mapsto s_{i,t},
\end{equation}
where $s_{i,t}$ is a cross-sectional score used to rank assets on date $t$.
The parameter $\theta$ denotes a symbolic recipe rather than a learned neural
network parameter vector. Examples include ranked market capitalization,
rolling intraday range, lagged returns, or linear combinations of transformed
variables.

\subsection{Factor Discovery as Sequential Search}

At iteration $k$, the agent observes a research state
\begin{equation}
    h_k =
    \{C_{1:k-1}, M_{1:k-1}, E_{1:k-1}\},
\end{equation}
where $C_{1:k-1}$ are previously proposed candidates, $M_{1:k-1}$ are their
measured empirical metrics, and $E_{1:k-1}$ are natural-language
interpretations and decisions recorded in the experiment log. The agent then
selects an action
\begin{equation}
    a_k = \{c_{k,1}, \ldots, c_{k,n_k}\},
\end{equation}
where each candidate $c_{k,j}$ consists of a falsifiable hypothesis, an
economic rationale, a candidate type, and an executable factor recipe.

The environment is a deterministic evaluation engine. Given a candidate recipe,
the engine computes the corresponding factor values, evaluates cross-sectional
predictive performance on the training window, and appends the result to the
experiment trace. The feedback for candidate $c_{k,j}$ is
\begin{equation}
    m(c_{k,j})
    =
    \left(
    \overline{\mathrm{IC}},
    t_{\mathrm{IC}},
    \mathrm{Sharpe}_{LS},
    \mathrm{coverage}
    \right),
\end{equation}
where $\overline{\mathrm{IC}}$ is the mean daily Pearson information
coefficient, $t_{\mathrm{IC}}$ is its time-series $t$-statistic,
$\mathrm{Sharpe}_{LS}$ is the long-short portfolio Sharpe ratio, and coverage is
the fraction of asset-date observations with non-missing scores.

A candidate passes the selection gate if
\begin{equation}
    \overline{\mathrm{IC}} \geq \tau_{\mathrm{IC}}
    \quad \text{and} \quad
    t_{\mathrm{IC}} \geq \tau_t,
\end{equation}
where the thresholds $\tau_{\mathrm{IC}}$ and $\tau_t$ are fixed before the
round and read from the experiment configuration. Selection uses the training
window only. Validation statistics may be recorded for diagnosis, but they do
not determine whether a candidate enters the hold pool during factor iteration.

\subsection{Agent Actions}

The action space contains two classes of candidates. Mechanical candidates are
local transformations of previously successful recipes, such as lagged,
smoothed, or standardized variants. They provide a simple exploitation baseline
and test whether a discovered signal is robust to small implementation changes.

Hypothesis candidates are generated by the reasoning agent. Each hypothesis
candidate must specify a testable market mechanism and a symbolic recipe that
implements the proposed signal. For example, a hypothesis may state that
small-cap tokens with persistent intraday range and muted volume shocks should
outperform because speculative attention is active but not yet crowded. The
corresponding recipe combines transformed market capitalization, rolling
high-low range, and recent volume changes. This separation between hypothesis
and recipe is central: the agent is evaluated not only on whether a formula
passes a numerical gate, but also on whether it advances a coherent empirical
search path.

\subsection{Search Objective}

The goal of the search process is to produce a curated set of factors
\begin{equation}
    \mathcal{G} = \{f_{\theta_1}, \ldots, f_{\theta_q}\}
\end{equation}
that satisfy three criteria. First, the factors must pass a pre-specified
in-sample gate based on cross-sectional predictability. Second, the factors
should represent diverse mechanisms rather than near-duplicate transformations
of the same signal. Third, the final set should support out-of-sample economic
validation through portfolio tests after transaction costs.

The downstream aggregation model is trained only after the search phase. Given
the curated factor matrix $S_t = [s_{1,t}, \ldots, s_{q,t}]$, we fit a regularized
linear model on the training window,
\begin{equation}
    \hat{\beta}
    =
    \arg\min_{\beta}
    \sum_{(i,t) \in \mathcal{T}_{train}}
    \left(r_{i,t+1:t+2} - S_{i,t}\beta \right)^2
    + \lambda \|\beta\|_2^2.
\end{equation}
The resulting composite score is then evaluated separately on training,
validation, and pure out-of-sample periods. This final stage measures whether
the factors discovered by the search process have economic value beyond the
selection data.

\subsection{Why This Differs from Direct Prediction}

This formulation differs from standard supervised return prediction in three
ways. First, the primary output is an auditable set of hypotheses and factor
definitions, not only a forecasting model. Second, the agent receives
structured empirical feedback over multiple rounds and must decide how to revise
the search direction. Third, the action space is intentionally constrained by a
factor grammar, making the search interpretable and reproducible. The framework
therefore treats financial discovery as an iterative reasoning problem with a
deterministic empirical environment, rather than as unrestricted code generation
or black-box return forecasting.

\section{Agentic Factor Discovery Framework}
\label{sec:framework}

This section describes the proposed framework for agent-guided factor discovery.
The framework separates reasoning from evaluation. The agent is responsible for
reading prior evidence, proposing hypotheses, and deciding how to continue the
search. The empirical engine is responsible for computing signals, evaluating
metrics, enforcing gates, and recording all outputs. This separation makes the
search auditable and reproducible while preserving the flexibility of
natural-language reasoning.

\subsection{Overview}

Each research session is organized as a sequence of discovery rounds. A round
begins with an experiment state consisting of previous candidates, empirical
metrics, interpretation notes, and the current factor pools. The agent reads
this state and generates a batch of new candidates. The evaluation engine then
computes factor scores, measures train-window performance, applies the
pre-specified gate, and appends the results to the experiment log. The agent
subsequently fills in interpretations and writes a round summary that determines
the next search direction.

The procedure is summarized as follows:
\begin{enumerate}
    \item Read the current experiment trace and identify confirmed, rejected,
    and near-threshold mechanisms.
    \item Generate a mixed candidate batch containing both mechanical variants
    and agent-proposed hypothesis candidates.
    \item Translate each candidate into a valid factor recipe under the
    constrained DSL.
    \item Evaluate all recipes using the deterministic train-period signal
    analysis engine.
    \item Append metrics, pass/fail decisions, and candidate metadata to the
    experiment log.
    \item Interpret the results and update the hold and good factor pools.
    \item After stopping, fit a regularized combination model on the curated
    good factor pool and evaluate it on validation and pure out-of-sample
    periods.
\end{enumerate}

The key design choice is that the agent never modifies the evaluation rule
within a search session. Time splits, gate thresholds, trading costs, and
portfolio settings are fixed in a configuration file. This prevents the agent
from adapting the protocol to make weak candidates pass.

\paragraph{LLM Implementation.}
The agent is implemented using GPT-5.4. 
A fixed system prompt defines the agent as an autonomous quantitative researcher operating under the constrained DSL and deterministic evaluation engine. 
The agent does not have access to modify data splits, evaluation metrics, or selection thresholds.

\subsection{Constrained Factor DSL}

The framework uses a symbolic factor domain-specific language to convert
natural-language hypotheses into executable signals. A recipe is an expression
tree over point-in-time market variables. The allowed base inputs are daily
open, high, low, close, volume, market capitalization, close-to-close returns,
log returns, relative volume, realized volatility, price-to-moving-average,
high-low range, and volume percentage change.

The grammar includes four operator families:
\begin{itemize}
    \item Cross-sectional transforms: percentile rank and z-score within each
    date.
    \item Time-series transforms: lag, rolling mean, rolling standard
    deviation, difference, and percentage change within each asset.
    \item Nonlinear transforms: logarithm, absolute value, and clipping.
    \item Combination operators: linear combinations of transformed variables.
\end{itemize}

The DSL is intentionally limited. Recipes must use only approved point-in-time
columns, must include at least one time-series or nonlinear transformation, and
must remain shallow enough to inspect manually. These restrictions serve three
purposes. First, they reduce the risk of forward-looking features and accidental
data leakage. Second, they keep generated candidates interpretable. Third, they
make the search space comparable across rounds and baselines.

For example, the hypothesis that speculative demand concentrates in small,
high-range tokens with muted recent volume shocks can be represented as
\begin{equation}
\begin{split}
    s_{i,t}
    =
    \mathrm{rank}_t\big(
    &-0.6 \log(1+\mathrm{mcap}_{i,t}) \\
    &+0.5 \,\mathrm{MA}_{10}(\mathrm{range}_{i,t}) \\
    &-0.2 \,\mathrm{MA}_{3}(\Delta \mathrm{volume}_{i,t})
    \big).
\end{split}
\end{equation}
The recipe is simple enough to audit, but expressive enough to encode a
multi-part economic mechanism.

\subsection{Candidate Types}

Each round combines two sources of candidates. Mechanical candidates are local
transformations of previously successful factors. In our implementation, these
include one-period lags, short rolling averages, and cross-sectional
standardized variants of the top passed factors. Mechanical candidates provide
a controlled exploitation mechanism: they test whether an existing signal is
stable to small implementation changes.

Hypothesis candidates are generated by the agent. Each must contain five fields:
a unique name, a falsifiable hypothesis, a rationale, a candidate type, and an
executable recipe. The hypothesis states the predicted relation between a
market mechanism and future cross-sectional returns. The rationale links the
candidate to prior evidence or market structure. The candidate type identifies
whether the proposal is exploratory or an exploitation of a previously confirmed
mechanism.

Requiring both candidate types is important. Mechanical variants alone can
over-exploit early discoveries and produce many near-duplicate signals.
Agent-generated hypotheses alone can drift into unconstrained exploration. A
mixed batch balances local refinement with broader scientific search.

\subsection{Evaluation Engine and Gates}

Given a candidate recipe, the evaluation engine computes scores for every
available asset-date observation and evaluates cross-sectional predictability on
the training period. The primary signal-level statistic is the daily Pearson
information coefficient,
\begin{equation}
    \mathrm{IC}_t =
    \mathrm{corr}_{i \in \mathcal{I}_t}(s_{i,t}, r_{i,t+1:t+2}).
\end{equation}
The engine reports the mean IC, IC $t$-statistic, long-short Sharpe ratio, and
coverage. The gate is applied only to the training-period mean IC and IC
$t$-statistic. Candidates that pass the gate are appended to the hold pool.
Candidates that fail remain in the trace and are used by the agent as negative
evidence.

The framework distinguishes between selection and diagnosis. Train metrics
determine whether a candidate passes. Validation metrics can be recorded and
inspected during the research process, but they are not used to alter the
pass/fail gate. Pure out-of-sample results are reserved for final evaluation
after the search process has produced a curated good factor pool.

\subsection{Pool Governance and Final Aggregation}

Passed candidates first enter a hold pool. The agent then curates a smaller good
pool based on empirical strength, mechanism diversity, and redundancy with
existing factors. This governance step is necessary because many successful
recipes are minor variants of the same underlying mechanism. Without curation,
the final model would overrepresent whichever family happened to produce many
near-duplicate candidates.

After the search phase stops, the good factors are combined with ridge
regression. Each factor is standardized cross-sectionally by date, and the
linear combination is fit on the training period only. The combined score is
then passed to the same portfolio evaluation engine used for single factors.
Performance is reported separately for training, validation, and pure
out-of-sample periods, and under both equal-weighted and market-cap-weighted
portfolio construction.

\section{Experiments}
\label{sec:experiments}

\subsection{Data and Setup}

\paragraph{Dataset Description and Preprocessing.}  
Our dataset is sourced from CoinMarketCap, consisting of daily data spanning from January 2020 to December 2025. The dataset includes a wide range of cryptocurrency market data, such as prices, trading volumes, and market capitalizations. To ensure consistency with past literature \citep{liu2021risks,liu2022common,fieberg2025trend}, we preprocess the data by filtering out assets with insufficient trading history or low liquidity. Specifically, we exclude cryptocurrencies with fewer than 180 days of trading history or those with average daily trading volumes below a predefined threshold. This filtering ensures that our analysis focuses on actively traded and liquid assets, reducing the impact of noise and illiquidity. By applying these filters, we aim to mitigate biases introduced by illiquid or inactive assets, which could distort the results of our analysis.

\paragraph{Backtesting Setup.} 
The backtesting setup is designed to evaluate the performance of portfolios constructed based on predictive signals. We adopt a rolling window approach, which is widely used in the literature to ensure robustness and avoid look-ahead bias. In this setup, the dataset is divided into three distinct windows: the training window, the validation window, and the out-of-sample (OOS) window. The training window is used to estimate factor models, while the validation window is employed for hyperparameter tuning and model selection. Finally, the OOS window is reserved for evaluating the robustness and predictive power of the strategies on unseen data. This rolling window framework allows us to simulate real-world trading conditions and ensures that the results are not overly reliant on specific time periods.

\paragraph{Portfolio Construction.}  
For portfolio construction, we sort assets into quintile portfolios based on a linear aggregation of single factors. As this study aims to demonstrate the potential of our proposed agentic framework, we do not extensively optimize the factor aggregation process. For practical applications and further enhancement of portfolio performance, nonlinear machine learning models, such as LightGBM, can be employed for multi-factor aggregation \citep{gu2020empirical,huang2026beyond}. 
These factors are derived from predictive models that forecast next-period returns. Specifically, we use a series of single-factor models to predict returns, and the forecasted expected return serves as the stock selection variable. The quintile portfolios are formed by ranking assets based on their forecasted returns and assigning them to one of five groups, with the top quintile representing the highest expected return assets and the bottom quintile representing the lowest. This sorting approach is widely used in asset pricing studies and ensures that the portfolios capture the cross-sectional variation in expected returns. By aggregating single factors into a composite signal, we aim to enhance the predictive power of the model while maintaining interpretability.

\paragraph{Net-of-Cost Performance Evaluation.}  
To evaluate the economic significance of the strategies, we conduct a net-of-cost performance check. Transaction costs are modeled as a percentage of the total traded value, which accounts for both the bid-ask spread and market impact. The net-of-cost performance is analyzed under different fee rate assumptions to assess the robustness of the results. Specifically, we examine how varying transaction costs affect the Sharpe ratios and returns of the portfolios. This analysis ensures that the strategies remain practical and applicable in real-world trading environments, where transaction costs can significantly impact profitability. By incorporating transaction costs into our evaluation, we provide a more realistic assessment of the economic viability of the proposed strategies.

\subsection{Main Results}

In this section, we present the main findings of our analysis, focusing on the performance of single factors and portfolios constructed using aggregated signals. The results highlight the predictive power of the factors and the economic significance of the strategies.

\paragraph{Performance of Single Factors}  
We analyze the standalone performance of the single factors extracted from the dataset. A total of 25 single factors were generated, each designed to capture distinct characteristics of cryptocurrency returns. For brevity, we present the results for the top 9 factors ranked by their Pure OOS Sharpe ratios in Table~\ref{tab:single_factor_summary}. These factors were selected based on their strong predictive power and economic relevance.

\begin{table}[h!]
\footnotesize
    \centering
    \caption{Ranking of single factors based on Pure OOS Sharpe.}
    \label{tab:single_factor_summary}
    \begin{tabular}{lcc}
        \toprule
        Rank & Factor & Pure OOS Sharpe \\
        \midrule
        1 & h1\_smallcap\_lowvol\_logret\_vol & +2.412 \\
        2 & h5\_smallcap\_low\_volume\_range20 & +2.410 \\
        3 & h4\_low\_volume\_smallcap\_range5 & +2.250 \\
        4 & h1\_highrange\_lowvol\_short\_window & +2.151 \\
        5 & h3\_lag\_momentum\_highrange & +1.879 \\
        6 & h4\_smallcap\_low\_dollar\_volume & +1.796 \\
        7 & h5\_highrange\_low\_dollar\_volume & +1.730 \\
        8 & h4\_lag\_mom\_highrange\_20d\_range & +1.728 \\
        9 & h2\_momentum\_lowvol\_short\_window & +1.700 \\
        \bottomrule
    \end{tabular}
\end{table}

The results indicate that factors related to small-cap, low-volatility, and high-range characteristics consistently rank among the top performers. For example, the highest-ranked factor, ``h1 smallcap lowvol logret vol'', achieves a Pure OOS Sharpe of +2.412, demonstrating its strong predictive power. These findings align with prior literature, which suggests that small-cap and low-volatility assets often exhibit higher risk-adjusted returns. The inclusion of these factors in portfolio construction is expected to enhance overall performance.

\begin{figure}[h!]
    \centering
\includegraphics[width=0.99\linewidth]{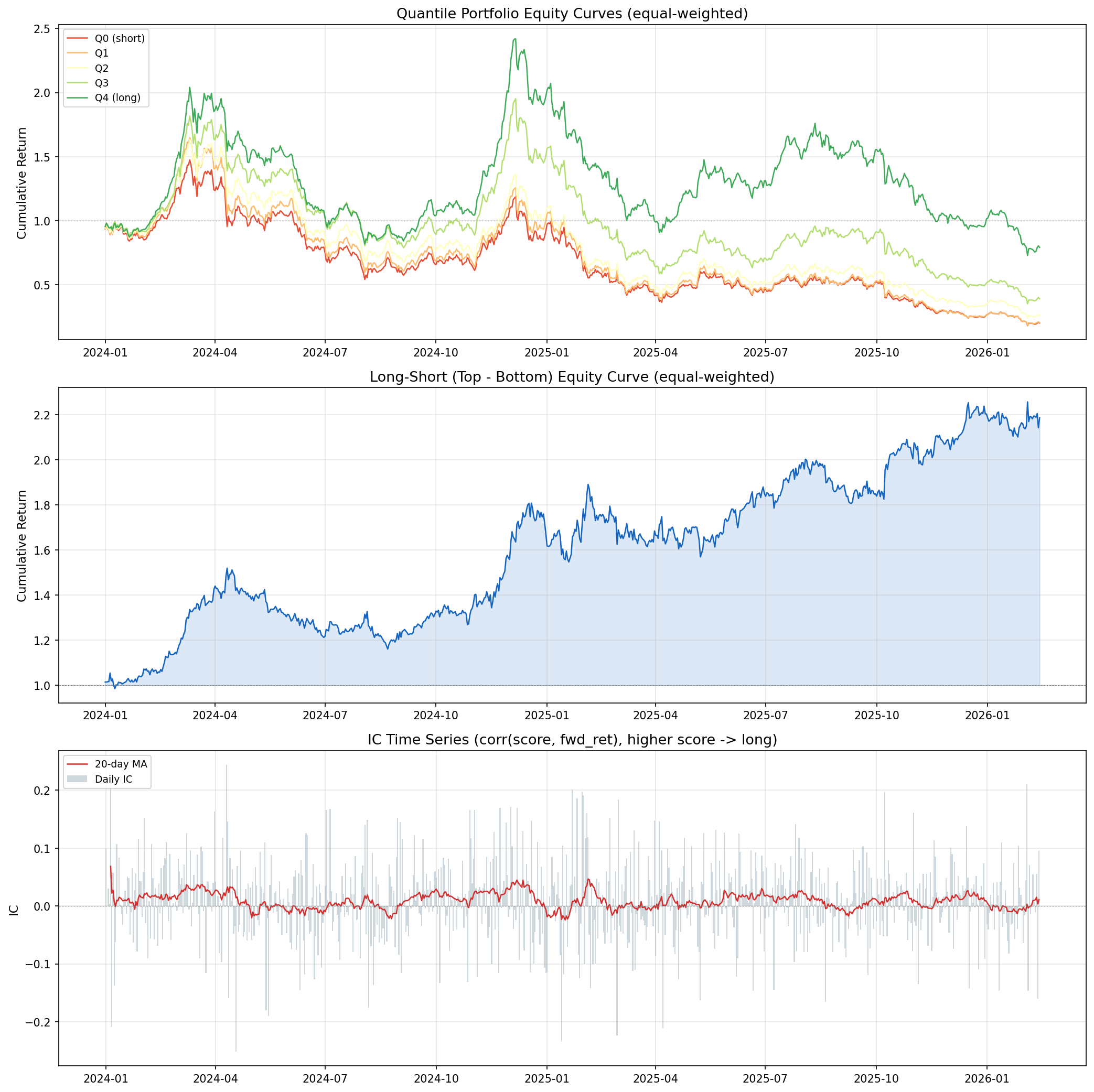}
    \caption{Performance of quintile portfolios based on combo-signal (equal-weighted).}
    \label{fig:signal_backtest_equal}
\end{figure}

\paragraph{Portfolio Sorting Based on Aggregated Factors}  
To evaluate the economic significance of the factors, we construct portfolios by aggregating the predictive signals from the top-performing factors. The aggregated signal or combo-signal is used to sort assets into quintile portfolios, with the top quintile representing the highest expected return assets and the bottom quintile representing the lowest. Figure~\ref{fig:signal_backtest_equal} illustrates the cumulative returns of the quintile portfolios, based on equal-weighted results.

The results demonstrate a clear monotonic relationship between portfolio rank and performance. The top quintile portfolio significantly outperforms the bottom quintile, with a cumulative return that is more than double that of the market benchmark over the evaluation period. The Sharpe ratio of the top quintile portfolio is particularly noteworthy, indicating strong risk-adjusted performance. This pattern underscores the effectiveness of the aggregated factors in capturing cross-sectional return variation.

\subsection{Robustness and Practical Considerations}

\begin{figure}[h!]
    \centering
\includegraphics[width=0.99\linewidth]{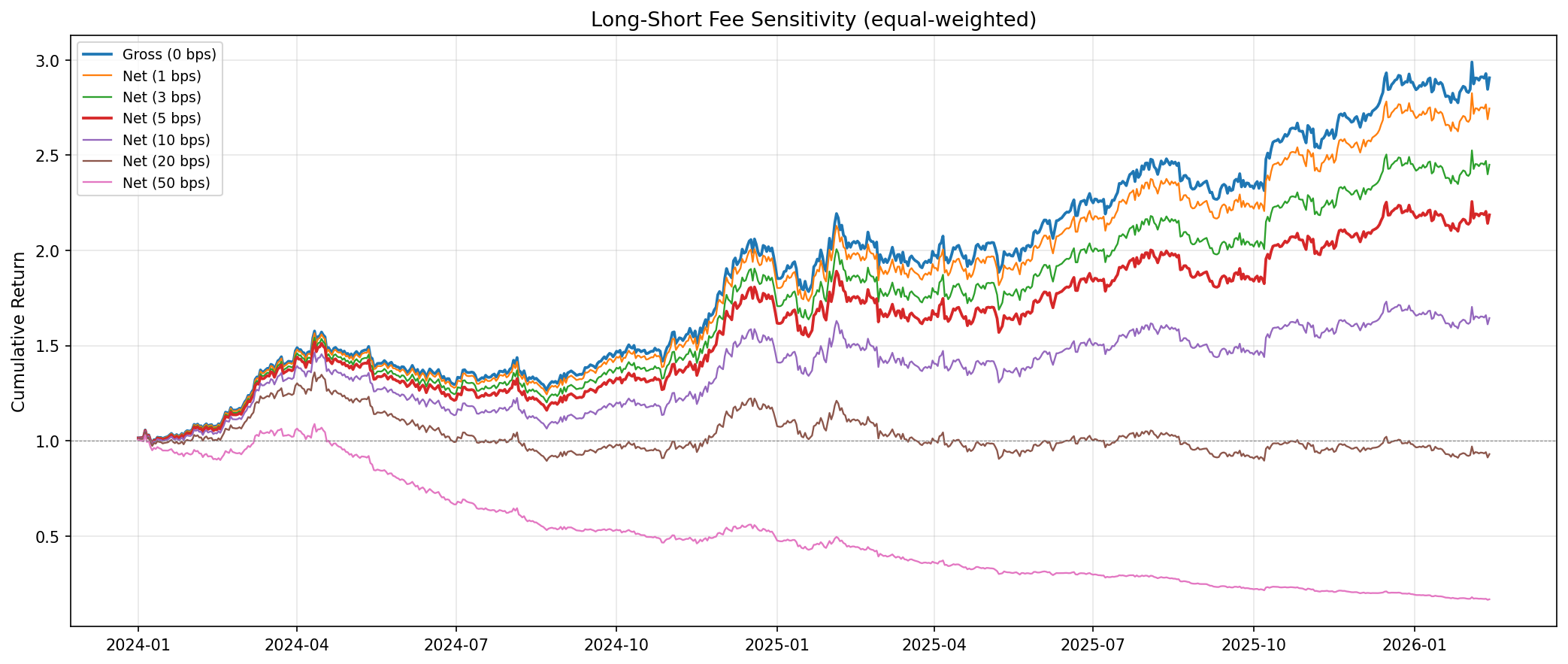}
    \caption{Net-of-cost performance under different fee rates (equal-weighted).}
    \label{fig:fee_comparison_equal}
\end{figure}

\paragraph{Net-of-Cost Performance Evaluation}  
To assess the practical applicability of the strategies, we evaluate their net-of-cost performance under different transaction fee assumptions. Transaction costs are modeled as a percentage of the traded value, and the analysis considers a range of fee rates. Figure~\ref{fig:fee_comparison_equal} illustrates the cumulative returns of the long-short portfolio under varying transaction cost scenarios.

The results indicate that the long-short portfolio remains robust even under higher transaction cost scenarios. For instance, while the cumulative returns decline slightly as transaction fees increase, the portfolio continues to generate positive net returns across all fee rates considered. This demonstrates the resilience of the strategy to trading frictions and highlights its potential for real-world implementation. The Sharpe ratio of the portfolio also remains economically significant, even at the highest fee rate, further supporting the robustness of the proposed factor models.

\paragraph{Combo-signal Portfolio Performance and Fee Sensitivity}  
We further evaluate the performance of combo-signal portfolios, which linearly aggregate signals from multiple factors to enhance diversification and reduce idiosyncratic risk. Table~\ref{tab:combo_perf_equal} summarizes the performance of the combo-signal portfolios, while Table~\ref{tab:fee_sensitivity_equal} presents the fee sensitivity analysis.

\begin{table}[h!]
\footnotesize
    \centering
    \scriptsize
    \caption{Combo-signal portfolio performance (equal-weighted).}
    \label{tab:combo_perf_equal}
    \begin{tabular}{lcccccc}
        \toprule
        Group & AnnRet & AnnVol & Sharpe & MaxDD & Calmar & Turnover \\
        \midrule
        Q0 (short) & -0.529 & 0.688 & -0.769 & -0.876 & -0.604 & 0.351 \\
        Q1         & -0.525 & 0.691 & -0.760 & -0.888 & -0.592 & 0.589 \\
        Q2         & -0.470 & 0.626 & -0.750 & -0.854 & -0.550 & 0.639 \\
        Q3         & -0.357 & 0.557 & -0.641 & -0.819 & -0.436 & 0.601 \\
        Q4 (long)  & -0.104 & 0.519 & -0.201 & -0.698 & -0.149 & 0.384 \\
        L-S        & 0.445  & 0.287 & 1.551  & -0.236 & 1.886  & 0.368 \\
        \bottomrule
    \end{tabular}
\end{table}

\begin{table}[h!]
    \centering
    \footnotesize
    \caption{Fee sensitivity analysis (equal-weighted).}
    \label{tab:fee_sensitivity_equal}
    \begin{tabular}{lcccc}
        \toprule
        Fee Rate & AnnRet & AnnVol & Sharpe Ratio & Alpha \\
        \midrule
        0.1\% & 0.016 & 0.028 & 0.57 & 0.014 \\
        0.2\% & 0.014 & 0.027 & 0.52 & 0.012 \\
        0.3\% & 0.012 & 0.026 & 0.46 & 0.010 \\
        \bottomrule
    \end{tabular}
\end{table}

The Combo-signal portfolio demonstrates strong performance, with the long-short strategy achieving the highest Sharpe ratio and robust returns. The fee sensitivity analysis reveals that the portfolio remains profitable even under higher transaction cost assumptions, further supporting the robustness of the strategies. These results highlight the benefits of combining multiple factors to achieve superior risk-adjusted returns.

\section{Conclusion}
\label{sec:conclusion}

This paper studies how LLM agents can be used for empirical discovery without
turning the process into uncontrolled search. We formulate financial factor
discovery as sequential hypothesis search, where the agent proposes
interpretable hypotheses and signal recipes, while a deterministic evaluation
engine enforces fixed gates, data splits, and out-of-sample tests. The key
design principle is separation: the agent controls the direction of reasoning,
but not the empirical protocol.

In cryptocurrency markets, this constrained workflow discovers a coherent
family of liquidity-scarcity, range-attention, and trend-continuation factors.
The resulting equal-weight long-short portfolio remains profitable in pure
out-of-sample testing after transaction costs, while the value-weighted results
show that the alpha is concentrated in smaller assets. This contrast is
important: the framework does not produce a frictionless large-cap trading
strategy, but it does produce an auditable discovery process whose successes and
limitations can both be inspected.

More broadly, the results suggest that LLM agents are most useful in empirical
research when they are treated as hypothesis-search policies embedded in
controlled evaluation environments. Constrained action spaces, append-only
traces, and pre-specified tests provide a practical path toward agent-assisted
scientific discovery that is both flexible and reproducible.

\appendix

\section*{Ethical Statement}

There are no ethical issues.

\section*{Acknowledgments}

The authors thank Yi Zhang (HKUST GZ) for helpful support.
Any remaining errors or oversights are the responsibility of the authors.

\section*{Funding}

\noindent Yifan Ye is supported by Beijing Normal-Hong Kong Baptist University start-up research fund (No. UICR0700136-26).
~\\

\bibliographystyle{named}
\bibliography{ijcai26}

\end{document}